\documentclass[12pt]{iopart}
\usepackage{iopams}  
\usepackage{graphicx}
\usepackage{dcolumn}
\usepackage{bm}
\usepackage{hyperref}
\usepackage{cite}

\usepackage{rotating}
\usepackage{array}
\usepackage{amssymb}
\usepackage{tikz}

\def\v#1{{\bf#1}}
\def\be{\begin{equation}}
\def\ee{\end{equation}}
\def\bea{\begin{eqnarray}}
\def\eea{\end{eqnarray}}

\newcommand{\bfalpha}{\mbox{\boldmath$\alpha$\unboldmath}}

\newcommand{\bfsigma}{\mbox{\boldmath$\sigma$\unboldmath}}
\newcommand{\bfpi}{\mbox{\boldmath$\pi$\unboldmath}}
\newcommand{\bfgamma}{\mbox{\boldmath$\gamma$\unboldmath}}

\newcommand{\bfkappa}{\mbox{\boldmath$\kappa$\unboldmath}}

\def\ccal{\mbox{$\cal C\,$}}

\def\ppcal{\mbox{$\cal P\,$}}
\def\tcal{\mbox{$\cal T\,$}}
\def\cliff{\mbox{$\mathcal{C}\ell$}}
\def\<{\langle}
\def\>{\rangle}

\begin{document}

\title[The stabilizer group of honeycomb lattices]{The stabilizer group of honeycomb lattices and its application to deformed monolayers}

\author{Y. Hern\'andez-Espinosa$^{1}$, A. S. Rosado$^{2}$ and E. Sadurn\'i$^{1, *}$}
\address{$^1$Instituto de F\'isica, Benem\'erita Universidad Aut\'onoma de Puebla,
Apartado Postal J-48, 72570 Puebla, M\'exico}
\address{$^2$Instituto de F\'{\i}sica, Universidad Nacional Aut\'onoma de M\'exico, 01000 M\'exico D.F., Mexico}
\ead{$^{*}$sadurni@ifuap.buap.mx}

\begin{abstract}
Isospectral transformations of exactly solvable models constitute a fruitful method for obtaining new structures with prescribed properties. In this paper we study the stability group of the Dirac algebra in honeycomb lattices representing graphene or boron nitride. New crystalline arrays with conical (Dirac) points are obtained; in particular, a model for dichalcogenide monolayers is proposed and analyzed. In our studies we encounter unitary and non-unitary transformations. We show that the
latter give rise to $\ppcal \tcal$-symmetric Hamiltonians, in compliance with known results in
the context of boosted Dirac equations. The results of the unitary part are applied
to the description of invariant bandgaps and dispersion relations in materials such as MoS$_2$. A careful construction based on atomic orbitals is proposed and the resulting dispersion relation is compared with previous results obtained through DFT. 
\end{abstract}





\section{Introduction}

A long-pursued goal in the study of crystals is to know whether an analogue of the Lorentz group exists in discrete space \cite{livine,baskal,georgieva}. While this fundamental question is of a mathematical nature, a strong physical motivation can be found in the happy analogy between the Dirac equation and honeycomb tight-binding models \cite{history,semenoff}. The study of electronic transport in graphene \cite{katsnelson,katsnelson2,semenoff2,ketnovo} and its emergent cousins \cite{novoselovcastro,balendhran,geimTMD} constitutes a relevant example. In fact, this motivation defines more clearly the mathematical problem of finding the invariance group of the Schr\"odinger equation for a hopping particle.

In this paper we deal with the problem of finding the action of the Lorentz group in effective Dirac theories by keeping a close look on physical concepts such as invariant dispersion relations and bandgaps in crystals, with applications to solid state physics and artificial lattice realizations in photonics \cite{chien}, phononics \cite{maynard}, microwaves \cite{sadurni,villasadurni} and cold atoms \cite{tarruell2012, lee2009} -- early realizations of artifical crystals with different purposes can be found, for example, in \cite{jaksch}. By means of a straightforward technique, we shall find the transformations of a nearest-neighbour Hamiltonian, giving rise to new operators whose spectrum is expressed with the same dispersion relation $E_k/\hbar = \omega$. More specifically, our goal is to show that the stabilizer group of the Dirac algebra (which contains Lorentz transformations) produces new lattices with the same bandgap.

More emergent applications of the Dirac equation can be found within the class of hexagonal semiconductors, in particular novel materials such as transition metal dichalcogenides \cite{mak2010, radisavljievic2011, duan2015}; see also \cite{kolobov2016} and references therein. Their peculiar structure shows the need of more complex tight-binding arrays that, however, lead to the typical bandgap emerging from inversion symmetry breaking precisely at a Dirac point. The question of whether such a description can be conceived as a unitary transformation of known models shall be addressed.

It is convenient to mention here that in our analysis, non-hermitian Hamiltonians shall emerge naturally. Although this is connected with similarity transformations that are non-unitary, the resulting non-hermitian operators will be proved to be $\ppcal \tcal$-symmetric \cite{bender,benderquantumtheory}, in compliance with real spectra. We recall here that a $\ppcal \tcal$-symmetric Hamiltonian has the remarkable property of commuting with the combination of parity and time reversal $\ppcal \tcal$, despite the fact that individually, neither parity nor time reversal commute with the Hamiltonian.

We proceed in three stages. First we pave the way with the one-dimensional case in section \ref{one}. We shall encounter compact and non-compact subgroups giving rise to unitary and non-unitary transformations. As a consequence, we shall classify our results in two sets: hermitian models that describe transport in slightly deformed solids and non-hermitian 
$\ppcal \tcal$-symmetric models that touch the realm of artificial realizations \cite{paritytime,longhi,bittner}. In section \ref{two} we deal with two-dimensional lattices, finding similar results. In section \ref{three} we discuss the plausibility of our new models in the context of dichalcogenide monolayers. A brief conclusion is given in section \ref{conclusion}.

\section{The Lorentz group applied to a one-dimensional lattice \label{one}}

Our task consists in expressing tight-binding models with Dirac points in terms of Minkowski vectors and Dirac $\gamma$ matrices.
The nearest-neighbour model of a particle in a linear chain of atoms or sites of two types, can be expressed as a Dirac Hamiltonian \cite{sadurni} employing suitable definitions of Dirac matrices in terms of localized (atomic) states $|n\>$. Denoting the nearest-neighbour hopping amplitude by $\Delta$ and the even (odd) site energy by $E_1$ ($E_2$) allows to write: 

\bea
H = \Delta \bfalpha \cdot \bfpi + \mu \beta + E_0, \quad \mu \equiv \frac{E_1-E_2}{2}, \quad E_0\equiv \frac{E_1+E_2}{2},
\label{1.1}
\eea

\bea
\fl \pi_1 \equiv 1+ \frac{1}{2} \sum_{n \in \mathbb{Z}} |n-2\>\<n | +  |n\>\<n-2 |, \quad
\pi_2 \equiv \frac{1}{2i} \sum_{n \in \mathbb{Z}} |n-2\>\<n | -  |n\>\<n-2|, 
\label{1.2}
\eea

\bea
\alpha_1 &\equiv&  \sum_{n \, \mbox{\scriptsize even}} |n+1\>\<n | +  |n\>\<n+1 |, \quad \nonumber \\ 
\alpha_2 &\equiv&  i \sum_{n \, \mbox{\scriptsize even}} |n+1\>\<n | -  |n\>\<n+1 |, \quad \nonumber \\
\beta &\equiv&  \sum_{n \, \mbox{\scriptsize even}} |n\>\<n |- |n+1\>\<n+1 |.
\label{1.3}
\eea
These definitions comply with the usual anticommutation relations $\left\{\alpha_i, \beta \right\}=0$ and satisfy the su$(2)$ algebra. Therefore, the matrices (\ref{1.3}) can be represented by Pauli matrices in the space of even and odd sites: $\alpha_1=\sigma_1$, $\alpha_2=\sigma_2$, $\beta=\sigma_3$. The 'kinetic' operators obey $\left[\pi_i, \pi_j\right]=0=\left[\sigma_i,\pi_j \right]$ and the Hamiltonian in (\ref{1.1}) has a full-band energy spectrum that depends crucially on the spectrum of $\bfpi$. The energies are

\bea
E^{\pm}_k = E_0 \pm \sqrt{4 \Delta^2 \cos^2 k + \mu^2 }, \qquad 0<k<\pi.
\label{1.4}
\eea
 It is worth mentioning that the usual $1+1$ Dirac equation is recovered when $k=k_D + \kappa$, $|\kappa| \ll k_D$ and $k_D$ is the Dirac point at $\pi/2$ (linearizing thus the cosine). In this limit, the eigenvalues of the periodic operators $\bfpi$ are such that $\pi_1 = 1 + \cos2k \rightarrow 0, \pi_2 = -\sin2k \rightarrow \kappa$. The Dirac $\gamma$ matrices can be obtained through $\gamma_0 = \sigma_3, \gamma_1 = \sigma_3 \sigma_1, \gamma_2 = \sigma_3 \sigma_2$, and it is straightforward to verify the corresponding algebra $\left\{ \gamma_{\mu}, \gamma_{\nu} \right\}=2 \eta_{\mu \nu}\v 1$. With these definitions and properties, one may write the Schr\"odinger equation ($\hbar=1$) $H\psi=i\partial\psi/\partial t$ in covariant form, i.e. multiplying both sides by $\gamma_0$ and subtracting the l.h.s. yields:

\bea
\left[\gamma_0 \left( i \frac{\partial}{ \partial t } - E_0 \right) - \Delta \bfgamma \cdot \bfpi - \mu \right] \psi = 0.
\label{1.5}
\eea
Furthermore, the substitutions $\Psi = e^{iE_0 t}\psi$, $\Pi^1 = \Delta \pi_1$, $\Pi^2 = \Delta \pi_2$, $\Pi^0 = i \partial/\partial t$ and the Einstein convention, lead to a compact and transparent expression

\bea
\left[\gamma_{\mu} \Pi^{\mu} - \mu \right] \Psi = 0, \quad \mu=0,1,2.
\label{1.6}
\eea
Evidently, the formal application of Lorentz transformations $\Lambda_{\mu \nu}$ on the objects $\Pi^{\mu}$ and $\gamma^{\mu}$ leave this equation invariant. The problem here is to provide a physical interpretation of such an invariance. We have the following observations:

\begin{itemize}
\item[i)] The transformation $\tilde{\Pi}^{\mu}=\Lambda^{\mu}_{\, \,\nu}\Pi^{\nu}$ changes $\Pi^0 \mapsto \tilde{\Pi}^0$; therefore, the Hamiltonian also suffers a transformation.
\item[ii)] The operators $\Pi^{\mu}$ satisfy an abelian algebra, but $\Pi^1, \Pi^2$ do not represent the generators of translations in space; therefore, there are no coordinates $X^{\mu}$ associated with Minkowski space \footnote{It has been proved \cite{sadurni2014, levi2001} that there exist canonically conjugate operators in crystals, similar to $x$ and $p$. However, if $X^{\mu}$ is defined as the canonical conjugate of $\Pi^{\mu}$ given by a combination of translations, the former will be non-local and its interpretation in terms of lattice deformations will be invalid.}.
\item[iii)] The dispersion relation (\ref{1.4}) is equivalent to $(E_k - E_0)^2 - (2\Delta \cos k)^2= \mu^2 $, or in operator form $\Pi_{\mu}\Pi^{\mu}\Psi=\mu^2\Psi$, which is manifestly an invariant. Therefore, any transformation $\Lambda_{\mu \nu}$ that preserves the Dirac (Clifford) algebra $\cliff (\v M_{2+1})$ produces a new equation with the same dispersion relation. In particular, the bandgap $\mu$ is an invariant -- sometimes this is identified with a fictitious rest mass. 
\end{itemize}
The invariance analysis must be focused on all the transformations preserving $\cliff (\v M_{2+1})$, i.e. a stabilizer. Since we are working with a specific $2\times2$ representation, we must have similarity transformations:

\bea
\tilde \gamma_{\mu} = S^{-1} \gamma_{\mu} S = \Lambda_{\mu}^{\,\, \nu} \gamma_{\nu}.
\label{1.7}
\eea
We see that the su$(2)$ algebra satisfied by (\ref{1.3}) is also preserved by (\ref{1.7}). In general $S\in$ GL$(2, \mathbb{C})$, but removing trivial scale factors leads to consider $S \in$ SL$(2,\mathbb{C})$, locally isomorphic to SO$(3,1)$. This set of transformations modify the Hamiltonian in the following way 

\bea
\tilde H = S^{-1} H S = \Delta \tilde{\bfgamma} \cdot \bfpi + \mu \tilde \beta + E_0, 
\label{1.8}
\eea
where we have used $\left[\bfpi,S\right]=0$. We divide our study in two classes of deformations: compact and non-compact subgroups.

\subsection{Compact subgroup}

\begin{figure}[t!]
\begin{center}  \includegraphics[width=\textwidth]{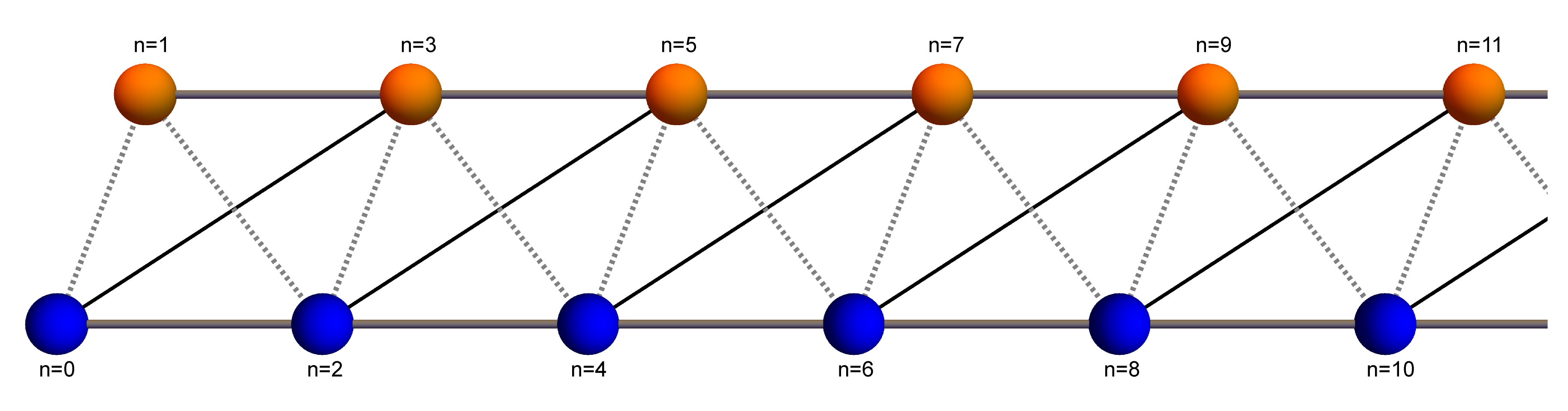} \end{center}
\caption{\label{fig:I.0} Emergent couplings in the lattice represented by (\ref{1.12}). Odd sites belong to the upper subchain (orange) and even sites to the lower subchain (blue). The lines represent couplings according to: nearest neighbours, dotted line; second neighbours, thick grey line; third neighbours, solid black line. The slanted configuration shows that third neighbours are coupled in the form $n, n+3$ with $n$ even.}
\end{figure}

%

Let us apply transformations in SO$(3)\subset$ SO$(3,1)$, comprising all rotations of the quasi-spin vector $\bfsigma$. The representations are unitary $S^{-1}=S^{\dagger}$. Crystal rotations in three (abstract) planes are available. Their meaning is better explained by the Foldy-Wouthuysen transformation \cite{devries}, which is a particular representative of rotations, turning $\tilde H$ into a diagonal operator in quasi-spin space and separating the upper and lower energy bands \footnote{This is the analogue of positive and negative energy states in the traditional Dirac theory, but our FW transformation considers the full energy band beyond conical points.}:

\bea
\fl \tilde H_{\scriptsize \mbox{FW}} &=& U^{\dagger} H U = \left( \begin{array}{cc} \sqrt{\Pi_1^2 + \Pi_2^2 + \mu^2} + E_0  & 0 \\  0 &  -\sqrt{\Pi_1^2 + \Pi_2^2 + \mu^2} + E_0 \end{array} \right),  \nonumber \\
\fl U&=&\exp \left( \frac{-i \phi \sigma_3}{2} \right) \exp \left( \frac{-i \theta \sigma_2}{2} \right),
\label{1.9}
\eea
where the angles $\theta, \phi$ commute with $H$, they are independent of $\bfsigma$, and are given by

\bea
\cos \theta = \frac{\mu}{\sqrt{\Pi_1^2 + \Pi_2^2 + \mu^2}}, \quad &\sin \theta& = \frac{\sqrt{\Pi_1^2 + \Pi_2^2}}{\sqrt{\Pi_1^2 + \Pi_2^2 + \mu^2}}, \nonumber \\
\cos \phi = \frac{\Pi_1}{\sqrt{\Pi_1^2 + \Pi_2^2}}, \quad &\sin \phi&= \frac{\Pi_2}{\sqrt{\Pi_1^2 + \Pi_2^2}}. 
\label{1.10}
\eea
A transformation $S = \exp (-i \v a \cdot \bfsigma)$ with arbitrary parameters generates other models. For rotations in the 1,2 plane we have $S= \exp (-i a_3 \sigma_3 )$ and the resulting Hamiltonian is merely an additional phase factor in couplings $\Delta \mapsto \Delta e^{\mp i a_3}$. In the case $S=\exp(-i a_1 \sigma_1)$ we have complex matrix elements as well. In order to obtain new real couplings, we study $S=\exp(-i a_2 \sigma_2)$, $\vartheta \equiv 2 a_2$. The effect is now visualized by means of (\ref{1.2}) and (\ref{1.3}) when replaced in the following expression  

\bea
\tilde H &=&  \Pi_1  \tilde \sigma_1 + \Pi_2 \tilde \sigma_2 + \mu  \tilde \sigma_3  + E_0 \nonumber \\
&=& \Pi_1 (\cos \vartheta \sigma_1 + \sin \vartheta \sigma_3) + \Pi_2 \sigma_2 + \mu (\cos \vartheta \sigma_3 - \sin \vartheta \sigma_1) + E_0 \nonumber \\
&=& H_1 + H_2 + H_3 + V
\label{1.11}
\eea
where $H_{1,2,3}$ contain couplings to first, second and third neighbours, and $V$ contains modified on-site energies:

\bea
\fl H_1 &=& \sum_{m \, \scriptsize \mbox{even}}  \frac{\Delta(\cos \vartheta + 1)}{2} |m-1\>\< m|  + \left( \Delta \cos \vartheta - \mu \sin \vartheta \right) |m+1\>\< m| + \mbox{h.c.}  \nonumber \\
\fl H_2 &=& \frac{\Delta \sin \vartheta}{2} \sum_{m \, \scriptsize \mbox{even}} | m+2 \>\< m| - |m+1 \>\< m-1| + \mbox{h.c.} \nonumber \\
\fl H_3 &=& \frac{\Delta(\cos \vartheta -1)}{2} \sum_{m \, \scriptsize \mbox{even}} | m+3 \>\< m | + \mbox{h.c.}                \nonumber \\
\fl V&=& E_0+ \left( \mu \cos \vartheta + \Delta \sin \vartheta \right)  \sum_{m \, \scriptsize \mbox{even}} | m \>\< m | -  | m+1 \>\< m+1 | .
\label{1.12}
\eea
These expressions display in each term, the necessary couplings to produce new models. Their shape is indicated in figure \ref{fig:I.0}, where bonds denote sites connected by hopping amplitudes. For slight deformations one has $\vartheta \ll 1$, showing that $H_2$ represents a linear correction, while $H_3$ is of second order and can be neglected.

\subsection{Non-compact subgroup and $\ppcal \tcal$-symmetry}

The transformations SO$(2,1) \subset$ SO$(3,1)$ rotate quasi-spin in the plane $(\sigma_1, \sigma_2)$ and produce boosts in the planes $(\gamma_0, \gamma_1)$ and  $(\gamma_0, \gamma_2)$. Let us analyze the hyperbolic transformations

\bea
\fl \left(\begin{array}{c} \tilde \gamma_0 \\ \tilde \gamma_2 \end{array} \right) = \left(\begin{array}{c} \exp(- \sigma_2 \varphi/2) \gamma_0 \exp( \sigma_2 \varphi/2)\\ \exp( -\sigma_2 \varphi/2) \gamma_1  \exp(\sigma_2 \varphi/2)  \end{array}\right) = \left(\begin{array}{cc} \cosh \varphi & \sinh \varphi \\ \sinh \varphi & \cosh \varphi \end{array} \right) \left(\begin{array}{c} \gamma_0 \\ \gamma_2 \end{array} \right),
\label{1.13}
\eea
which produce $\tilde \sigma_3 = \cosh \varphi \, \sigma_3 - i \sinh \varphi \, \sigma_1, \tilde \sigma_1 = \cosh \varphi \, \sigma_1 + i \sinh\varphi \, \sigma_3, \tilde \sigma_2 = \sigma_2$. The Hamiltonian becomes 

\bea
\tilde H &=& \Pi_1 (\cosh \varphi \sigma_1 + i \sinh \varphi \sigma_3) + \Pi_2 \sigma_2 + \mu (\cosh \varphi \sigma_3 - i\sinh \varphi \sigma_1) + E_0 \nonumber \\
&=& H_1 + H_2 + H_3 + V.
\label{1.14}
\eea
As before, we have up to third neighbour couplings in the expressions

\bea
 H_1 &=& \sum_{m \, \scriptsize \mbox{even}}  \frac{\Delta(\cosh \varphi + 1)}{2} \left\{ |m-1\>\< m|  + |m\>\< m-1| \right\} \nonumber \\ 
 &+& \left( \Delta \cosh \varphi - i \mu \sinh \varphi \right)\left\{|m+1\>\< m| + |m\>\< m+1| \right\}  \nonumber \\
 H_2 &=& \frac{i \Delta \sinh \varphi}{2} \sum_{m \, \scriptsize \mbox{even}}\left\{ | m+2 \>\< m| - |m+1 \>\< m-1| \right\} - \mbox{h.c.} \nonumber \\
 H_3 &=& \frac{\Delta(\cosh \varphi -1)}{2} \sum_{m \, \scriptsize \mbox{even}} | m+3 \>\< m | + \mbox{h.c.}                \nonumber \\
 V&=& E_0+ \left( \mu \cosh \varphi + i \Delta \sinh \varphi \right)  \sum_{m \, \scriptsize \mbox{even}} | m \>\< m | -  | m+1 \>\< m+1 | .
\label{1.15}
\eea
Similar observations on the structure in figure \ref{fig:I.0} apply to this result, except for the presence of skew-hermitian terms in $V, H_1, H_2$. Evidently, the lack of unitarity implies $\tilde H^{\dagger} \neq \tilde H$, but the spectrum remains real if we use a Hilbert space with a metric

\bea
\< k,s| S^{\dagger}S  |k',s' \> = \delta(k-k') \delta_{s,s'}, \quad s= \pm, \quad  0<k<\pi,
\label{1.16}
\eea
where $s=\pm$ is the upper and lower band index. Moreover, $\tilde H$ can be shown to be $\ppcal \tcal$-symmetric with a suitable definition of parity $\ppcal$. We reverse the signs in the 1,2 plane $\ppcal (\sigma_1,\sigma_2,\sigma_3) = (-\sigma_1,-\sigma_2,\sigma_3)$ and $\ppcal (\Pi_1,\Pi_2)=(-\Pi_1,-\Pi_2)$. From the definitions (\ref{1.1}), (\ref{1.2}) and (\ref{1.3}), the complex conjugation operator leaves $H$ invariant, i.e. $H^*=H$ and the overall antiunitary operation is

\bea
\fl \ppcal \tcal \left[H(\bfpi) \right] = \ppcal \left[H^*(\bfpi)\right] = \sigma_3 H^*(-\bfpi) \sigma_3 =  \sigma_3 H(-\bfpi) \sigma_3 =  H(\bfpi).
\label{1.17}
\eea
Now, with the help of $\sigma_2^* = -\sigma_2, S^* = S^{-1}$ and $\sigma_3 S^{-1} = S \sigma_3 $, the invariance of $\tilde H$ follows

\bea
\fl \ppcal \tcal \left[ \tilde H(\bfpi) \right] = \ppcal \left[ S H^*(\bfpi) S^{-1} \right] = \sigma_3 S H^*(-\bfpi) S^{-1} \sigma_3 = S^{-1} \sigma_3 H(-\bfpi) \sigma_3 S =  \tilde H(\bfpi).
\label{1.18}
\eea
It is also possible to identify the transformation $H \mapsto \tilde H$ with the addition of a $\ppcal  \tcal$-symmetric (but non-hermitian) potential

\bea
\tilde H = H + S^{-1} \left[ H,  S \right] \equiv H + U, \quad U^{\dagger}\neq U, \quad \ppcal \tcal \left[ U \right]=U,
\label{1.19}
\eea
where the transformation properties can be checked also infinitesimally $U \approx \varphi \left[ H, \sigma_2 \right]$. The appearance of $\ppcal \tcal -$symmetry in connection with boosts was noted before in \cite{benderfasterthan,brachistochrone}. We have shown here how to build a lattice that enjoys such property. Related work in photonic lattices can be found in \cite{szameit2011}, while the extended discrete symmetry $\ccal \ppcal \tcal$ was studied in \cite{yesiltas2013}, with additional contributions to symmetry breaking in \cite{sadurni2015}.

\begin{figure}[t!]
\begin{center}  \includegraphics[width=\textwidth]{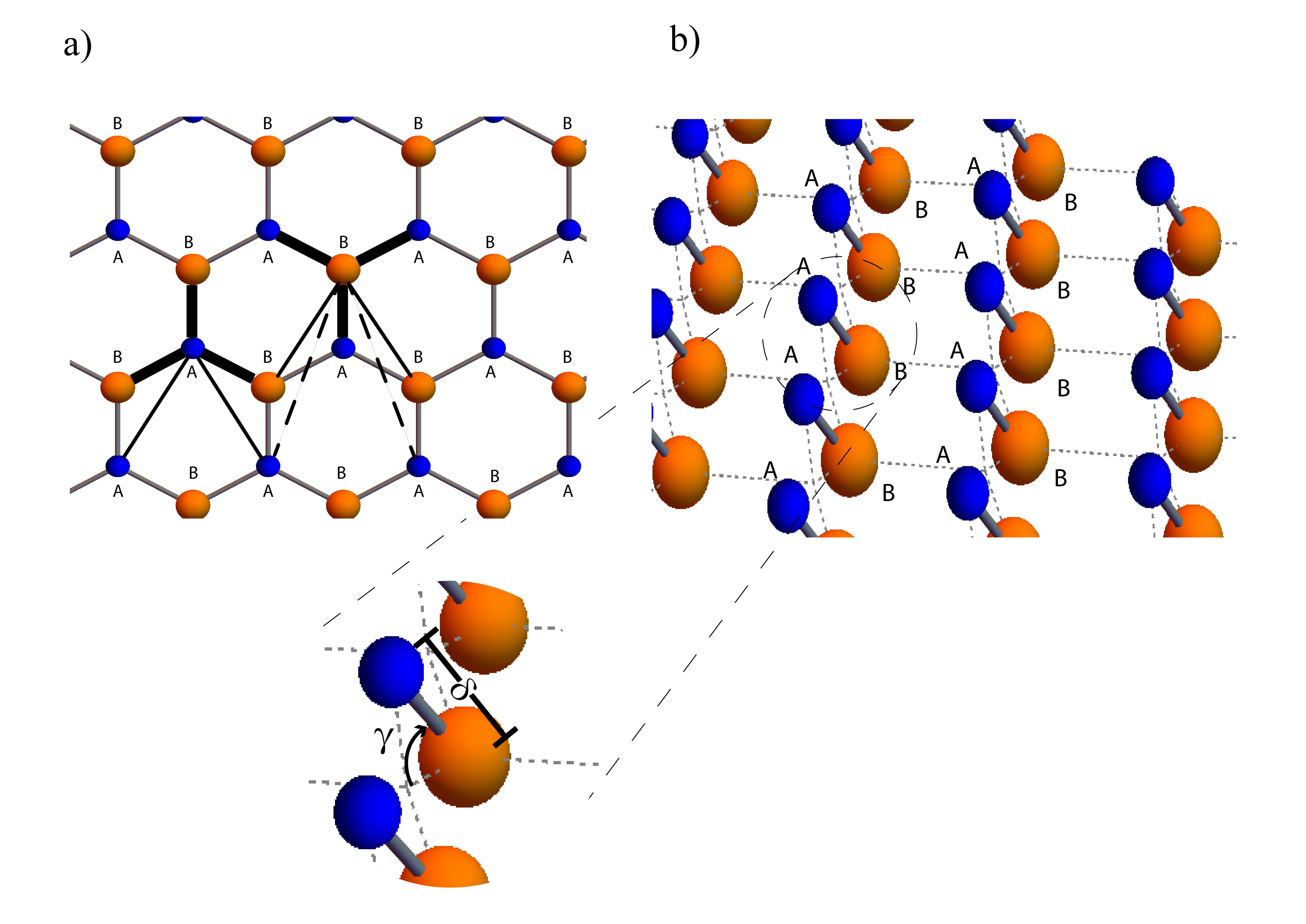} \end{center}
\caption{\label{fig:II.0} Emergent couplings in honeycomb structures. a) Thick black lines: nearest neighbours. Thin black lines: second neighbours. Dashed lines: third neighbours. b) Geometrically deformed lattice, allowing couplings by proximity. Since localized wavefunctions decrease exponentially with distance,  the couplings given by overlaps (\ref{3.2}, \ref{3.3}) decay with the neighbour range, i.e. $\<H_1\> > \<H_2\> > \<H_3\>$. Hence the appearance of deformation parameters $\gamma, \delta$, in compliance with (\ref{2.4}). }
\end{figure}

\section{The Lorentz group applied to honeycomb lattices \label{two}}

This is the most interesting case in terms of possible applications. Here we recall that the bipartite structure of the hexagonal lattice ensures the existence of Dirac matrices in $2\times2$ representation, which are defined exclusively in terms of localized states:

\bea
\Pi_1 \equiv \frac{\Delta}{2} \sum_{\v A, i=1,2,3} \left\{|\v A\>\< \v A + \v b_i - \v b_1 | + |\v A + \v b_1\>\< \v A + \v b_i | \right\} + \mbox{h.c.}, \nonumber \\
 \Pi_2 \equiv \frac{i\Delta}{2} \sum_{\v A, i=1,2,3} \left\{|\v A\>\< \v A + \v b_i - \v b_1 | + |\v A + \v b_1\>\< \v A + \v b_i | \right\} + \mbox{h.c.},
\label{2.1}
\eea

\bea
\alpha_1 &\equiv&  \sum_{\v A} |\v A+\v b_1\>\<\v A | +  |\v A\>\<\v A+\v b_1 |, \quad \nonumber \\ 
\alpha_2 &\equiv&  i \sum_{\v A} |\v A+\v b_1\>\<\v A | -  |\v A\>\<\v A+\v b_1 |, \quad \nonumber \\
\beta &\equiv&  \sum_{\v A} |\v A\>\<\v A |- |\v A+\v b_1\>\<\v A+\v b_1 |,
\label{2.2}
\eea
and the primitive vectors with unit lattice spacing are such that
\bea
\v A = n_1 \v a_1 + n_2 \v a_2, \quad \v a_1 = \v b_3-\v b_2, \quad \v a_2 = \v b_1-\v b_3 \nonumber \\
\v b_1 = \v j, \quad \v b_2 = -\frac{\sqrt{3}}{2} \v i - \frac{1}{2} \v j, \quad \v b_3 = -\v b_1 - \v b_2.   
\label{2.3}
\eea
The dispersion relation is well known: 
\bea
E_{\v k}^{\pm}=E_0 \pm \sqrt{\Delta^2|\sum_i e^{i\v k \cdot \v b_i}|^2+\mu^2}. 
\label{2.3.1}
\eea
It is important to mention that, in contrast with the $1+1$ dimensional case, around the (inequivalent) Dirac points $\v k_D = (4\pi/3\sqrt{3}) \v i , (2\pi/3\sqrt{3}) \v i +(2\pi/\sqrt{3}) \v j$ there is a linearization of both operators $\pi_1, \pi_2$ in terms of $\bfkappa = \v k_D - \v k$, where $\v k$ is the Bloch vector. When it comes to the application of the Lorentz group, the same reasoning as in the $1+1$ case applies here. Once more we have a Dirac algebra $\left\{ \gamma_{\mu} \right\}, \mu =0,1,2$ represented by Pauli matrices and the stabilizer is SL$(2, \mathbb{C})$. The honeycomb analogues of (\ref{1.12}) and (\ref{1.15}) are, respectively

\bea
\fl H_1 &=& \frac{\Delta(\cos \vartheta +1)}{2} \sum_{\v A, i=1,2} |\v A + \v b_i \>\<\v A| + \left( \Delta \cos \vartheta - \mu \sin \vartheta \right) \sum_{\v A} |\v A + \v b_1\>\<\v A| + \mbox{h.c.} \nonumber \\
\fl H_2 &=& \frac{\Delta \sin \vartheta}{2} \sum_{\v A, i=1,2} \left\{|\v A \>\<\v A + \v b_1 - \v b_i| - |\v A + \v b_i \>\<\v A + \v b_1|\right\} + \mbox{h.c.} \nonumber \\
\fl H_3 &=&  \frac{\Delta(\cos \vartheta -1)}{2}  \sum_{\v A, i=1,2} |\v A + \v b_1 \>\<\v A + \v b_i - \v b_1| + \mbox{h.c.}  \nonumber \\
\fl V &=& \left( \mu \cos \vartheta + \frac{\Delta \sin \vartheta}{2} \right) \sum_{\v A} \left\{ |\v A\>\<\v A| - |\v A + \v b_1\>\<\v A + \v b_1| \right\} + E_0,
\label{2.4}
\eea
and (replacing $\vartheta \mapsto i\varphi$ in $S$)

\bea
\fl H_1 &=& \frac{\Delta(\cosh \varphi +1)}{2} \sum_{\v A, i=1,2} \left\{ |\v A + \v b_i \>\<\v A| + |\v A  \>\<\v A + \v b_i|\right\} \nonumber \\ \fl &+& \left( \Delta \cosh \varphi - i\mu \sinh \varphi \right) \sum_{\v A}\left\{ |\v A + \v b_1\>\<\v A| + |\v A \>\<\v A + \v b_1 | \right\} \nonumber \\
\fl H_2 &=& i\frac{\Delta \sinh \varphi}{2} \sum_{\v A, i=1,2} \left\{|\v A \>\<\v A + \v b_1 - \v b_i| - |\v A + \v b_i \>\<\v A + \v b_1|\right\} - \mbox{h.c.} \nonumber \\
\fl H_3 &=&  \frac{\Delta(\cosh \varphi -1)}{2}  \sum_{\v A, i=1,2} |\v A + \v b_1 \>\<\v A + \v b_i - \v b_1| + \mbox{h.c.}  \nonumber \\
\fl V &=& \left( \mu \cosh \varphi + i \frac{\Delta \sinh \varphi}{2} \right) \sum_{\v A} \left( |\v A\>\<\v A| - |\v A + \v b_1\>\<\v A + \v b_1| \right) + E_0,
\label{2.5}
\eea
where the calculations involve $\<\v A | \v A' \>=\delta_{\v A, \v A'}$ and a careful redefinition of summation indices (valid only for infinite sheets). The new couplings give rise to new links representing hopping amplitudes, depicted in figure \ref{fig:II.0}.

\begin{figure}[t!]
\begin{center}  \includegraphics[width=\textwidth]{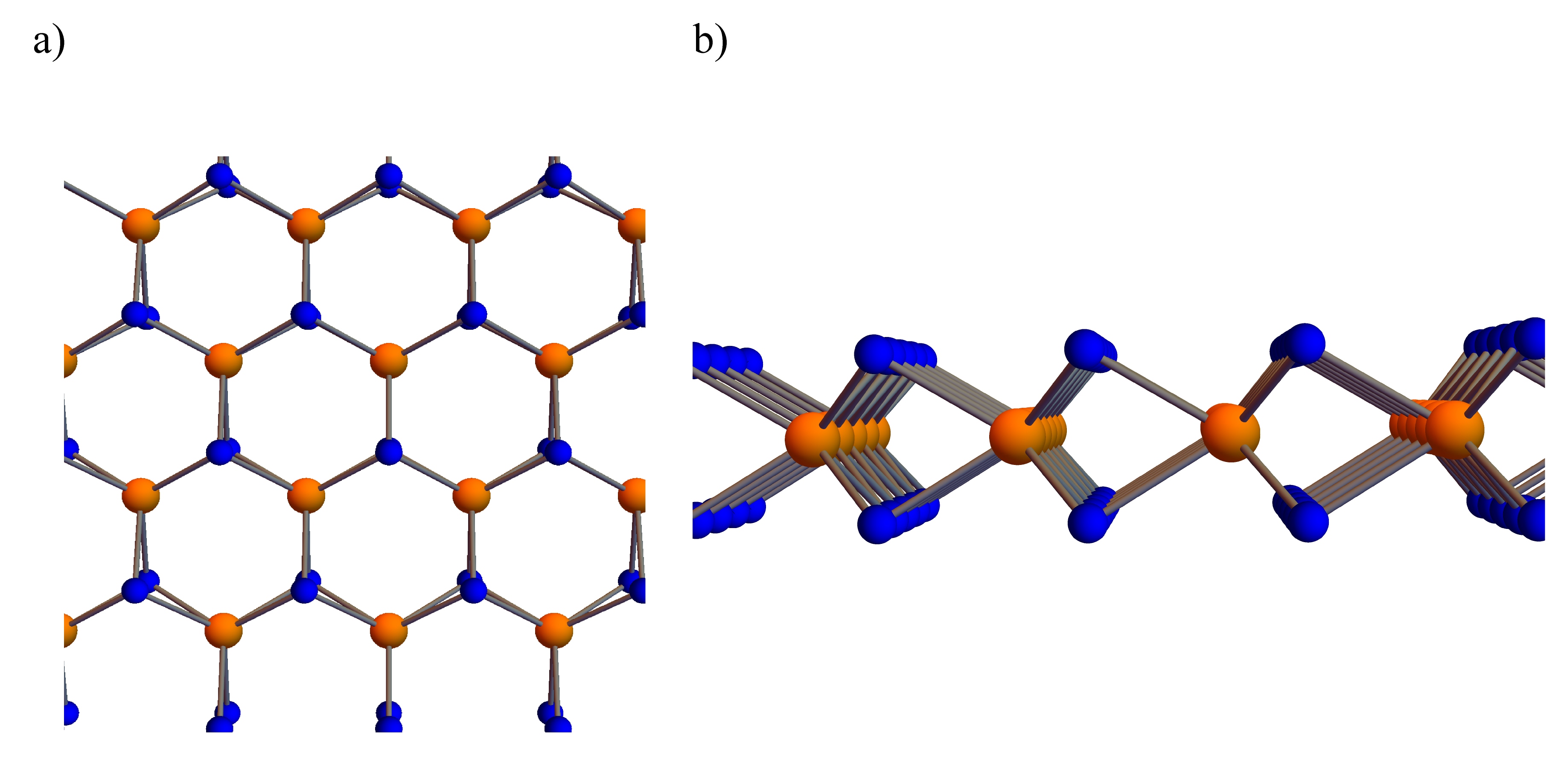} \end{center}
\caption{\label{fig:III.0} Crystalline structure of MoS$_2$. a) Honeycomb lattice. b) Buckled structure exposed in a lateral view. The couplings by proximity are again justified as in fig. \ref{fig:II.0}, but now the blue sites in upper and lower layers (S centres) contain the symmetric wavefunctions in the first line of (\ref{3.1}) as a single state $|\v A\>$.}
\end{figure}

\begin{figure}[t!]
\begin{center}  \includegraphics[width=\textwidth]{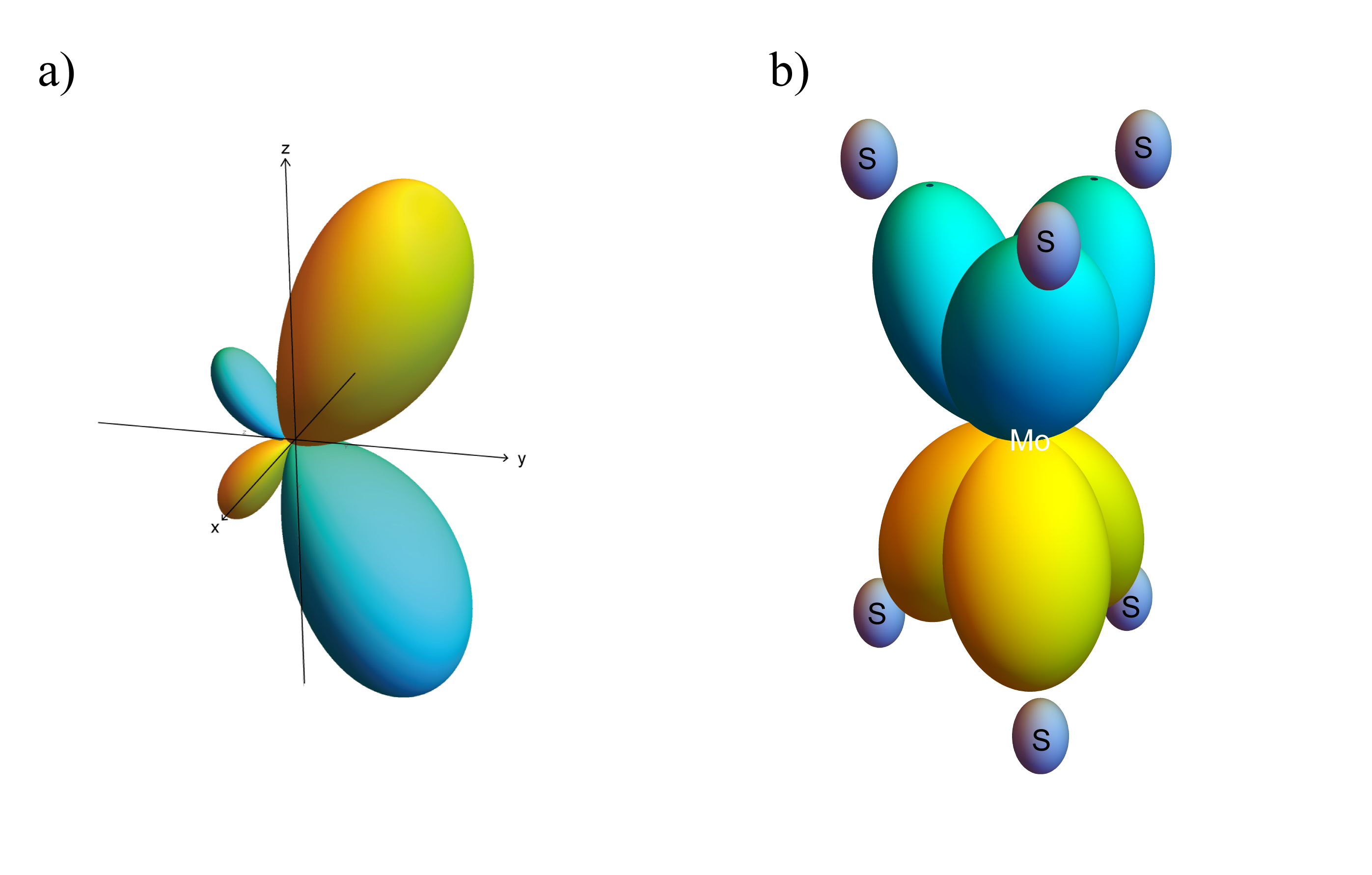} \end{center}
\caption{\label{fig:IV.0} Mo hybrid orbitals forming chemical bonds. a) A combination $ h_1 p_z + h_2 d_{xz} $ with $h_1 = 0.28, h_2 = 0.96$ has been used to produce the necessary angle for covalent bonds. b) Full ligands obtained by two $2\pi/3$ rotations of panel a).  }
\end{figure}

\begin{figure}[t!]
\begin{center}  \includegraphics[width=\textwidth]{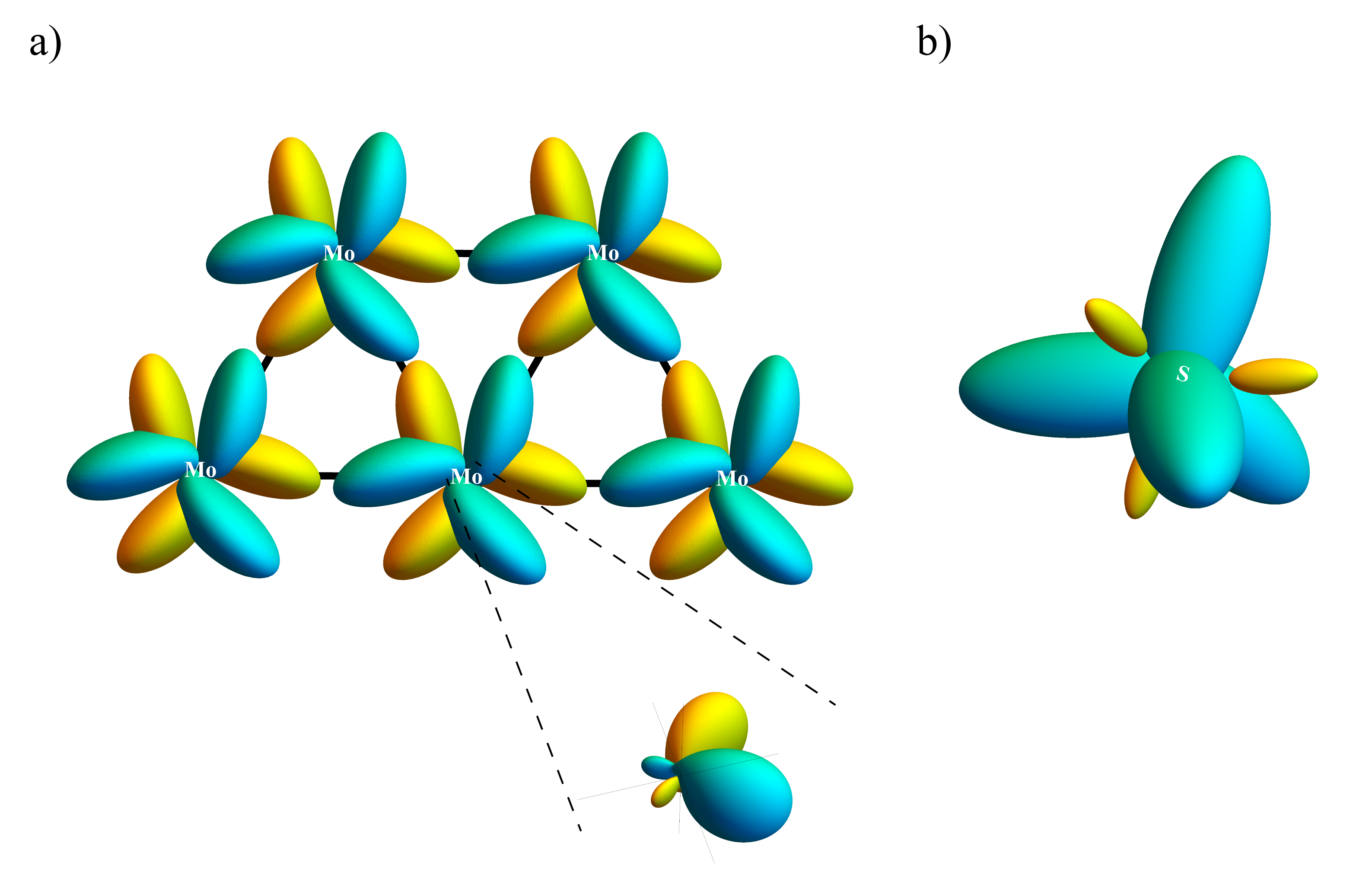} \end{center}
\caption{\label{fig:V.0} Orbitals used for conduction. a) Mo orbitals producing second neighbour interaction. Blue and gold represent plus and minus signs of wavefunctions, respectively. This construction allows negative couplings (blue meets gold), as described by $H_2$ in (\ref{2.4}). In the inset we see the hybrid orbital in the second line of (\ref{3.1}), before rotation. b) Typical tetrahedral structure of chalcogenides. The conduction orbital points upwards. }
\end{figure}

\section{Bandgaps in dichalcogenide monolayers \label{three}}

A close resemblance to Silicene, Germanene and Stanene \cite{balendhran} is observed in the emerging lattices shown in fig. \ref{fig:II.0}. This would be a good microscopic model for such materials, except for the opposite signs of second neighbour couplings, which are rather artificial for monoatomic sheets. We look now for a plausible scenario in which our second neighbour model -- with $\vartheta < 1 $ and $\< H_3 \> \sim 0$ -- can describe a well-known semiconductor structure made of two species (thus the second neighbour couplings may differ). Dichalcogenides are good candidates; in particular MoS$_2$ (synthesized and characterized \cite{zhan,kin}) for which numerical studies have shown its semiconductor (direct) gap \cite{ridolfi,liu,rostami}, its full bandwidth and the range of validity in $k$ space based on the regularity of the band; also numerically-obtained nearest neighbour models have been proposed \cite{zahid} to explain these properties. The latter, although exact and carefully built, do not provide an intuitive construction based on overlaps between neighbouring sites, nor a simple mechanism to find the numerical parameters of the model. Our task is now to describe such systems in a simple manner, including second neighbours (\ref{2.4}), which leads by construction to the desired dispersion relation (\ref{2.3.1}). The first difficulty we encounter is related to the atomic orbitals. The monolayer material (X-M-X with one metallic and two chalcogen layers) must be a covalent crystal (as opposed to ionic), which dictates the nature of the overlaps that constitute chemical bonds. We must first discriminate such states in order to work with the remaining orbitals as candidates for conduction. In MoS$_2$, each Mo centre has six chemical  links to the nearest S centres. The six available orbitals in Mo cannot be used to form the six chemical bonds, since all orbitals would then be occupied, ruling out any hopping amplitude\footnote{This configuration would produce conduction only through second neighbours, i.e. between S centres disposed in a triangular lattice. The resulting band structure would differ considerably from the one observed.} between S and Mo. Therefore, one must find  less than six wave functions with a lobular structure resembling the buckled configuration in fig. \ref{fig:III.0}. This is possible if we resort to the hybridization of the $pd$ orbitals in the $n=4$ shell of Mo, i.e. linear combinations of $l=1$ and $l=2$. First, we build a wavefunction that locks one Mo atom to two S atoms, one in the upper and one in the lower layer. A combination of $p_z$ and $d_{xz}$ does the job, as shown in fig. \ref{fig:IV.0}. Then, the other two bonds can be obtained by two $2\pi/3$ rotations (if the lobes are sufficiently elongated, these three hybridized orbitals will be orthogonal). Filling these orbitals with three electrons reduces the number to three available orbitals for conduction. The shape of conduction orbitals must be such that their overlap with the hybridized $sp_z$ of S (pointing out of the sheet) must be the largest coupling in the model. The overlap of these Mo orbitals with their six second neighbours must have opposite sign: the product of adjacent wavefunctions must be negative, as per result (\ref{2.4}) second term of $H_2$, and there must be six lobes horizontally disposed as shown in fig. \ref{fig:V.0} a). This can be achieved by considering a linear combination of $p_x$ and $d_{xy}$ (available), together with a $p_z$ contribution that couples with S. This is shown in fig. \ref{fig:V.0} b). The full structure is built again by rotating $2\pi/3$ twice, giving rise to six horizontal leaves of alternating signs.

\begin{figure}[t!]
\begin{center}  \includegraphics[width=9cm]{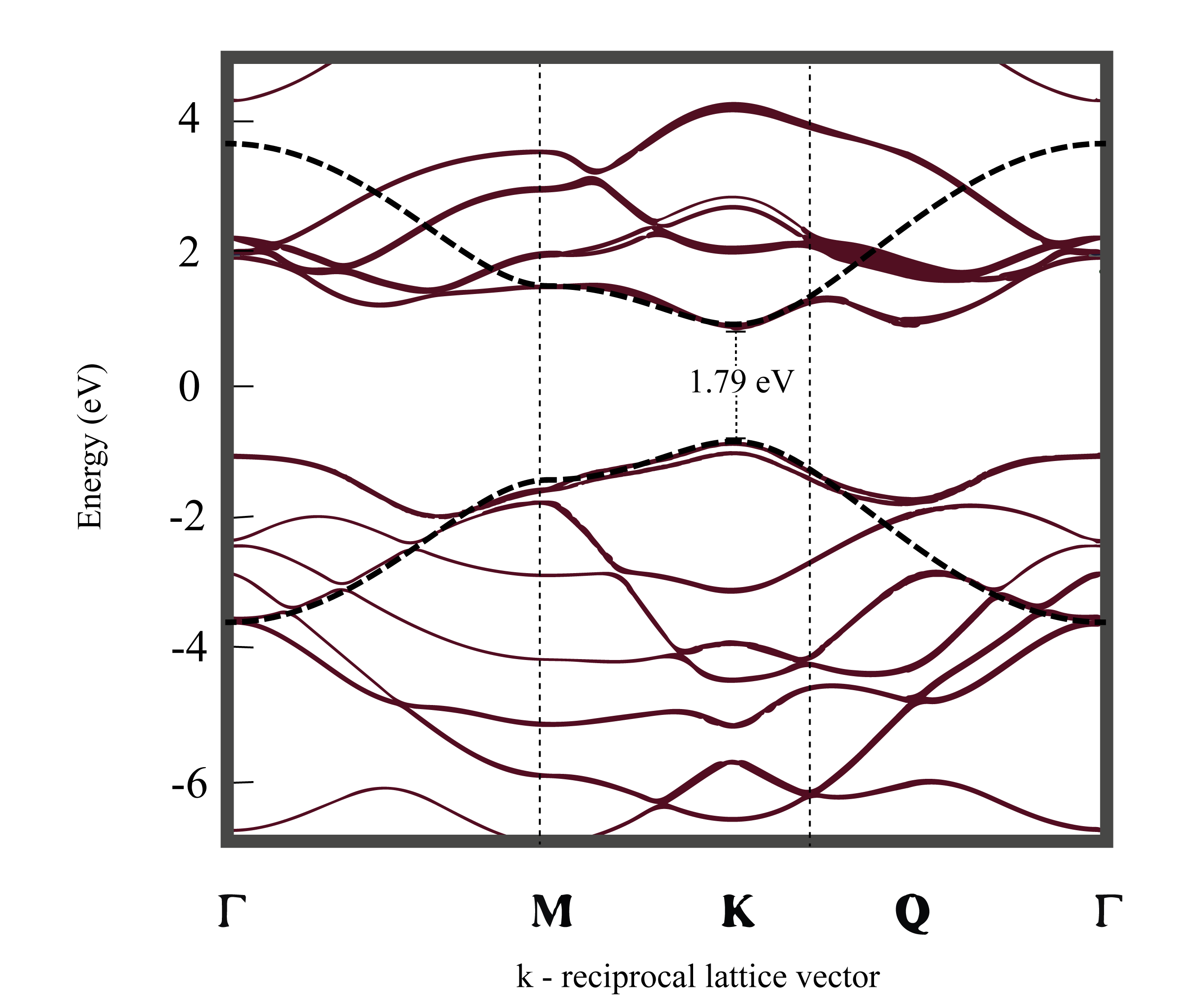} \end{center}
\caption{\label{fig:VI} Comparison between numerical bands (red, solid curve) and our simplified model (thick, dashed curve). The numerical value of the gap 1.79 eV has been introduced in our formula (\ref{2.3.1}) for a better comparison, but the experimental value 1.9 eV can be used as well. The vertical dashed lines indicate the limits of the region where the comparison is valid; our model does not describe indirect processes associated with the point Q. Remarkably, outside the region of interest and for the lower band, the black curve also reproduces the level (anti) crossings.}
\end{figure}

We have

\bea
\< \v r | \v A \> = \frac{1}{\sqrt{2}} ( sp_{z \, \scriptsize \mbox{lower}} + sp_{z \, \scriptsize  \mbox{upper}} )  \nonumber \\
\< \v r | \v B \> =  \left[ \v 1 + \v D_z(2\pi/3) + \v D_z(4\pi/3) \right]\left[c_1 p_{x} + c_2 d_{xy} + c_3 p_{z} \right]
\label{3.1}
\eea
where $c_i$ are real constants, $\v D_z$ are Wigner rotations around $z$, $\v B$ are Mo centres, $\v A$ are S centres and a symmetric combination of up and down chalcogen layers has been used as a single site (in our model, there is no coupling between these layers). The angle of deformation can be finally determined in terms of atomic overlaps by 

\bea
\fl \cos \vartheta &=& \frac{2 |\<\v A + \v b_2 | H | \v A \>|}{ \Delta} - 1 
= \frac{4 |\int \<\v r|\v B \> \left[ \frac{p^2}{2m} + V_{\scriptsize \mbox{crystal}} \right] \< \v A| \v r \>  dV  | }{\sqrt{E^2_{\scriptsize \mbox{bandwidth}}-E^2_{\scriptsize \mbox{gap}}}}-1
\label{3.2}
\eea
where $E_{\scriptsize \mbox{bandwidth}} \approx 2 \sqrt{\Delta^2+ \mu^2}, E_{\scriptsize \mbox{gap}}\approx 2\mu$ are realistic quantities approximated by our model parameters. We may also eliminate phenomenological parameters to extract the angle in terms of nearest neighbours (denominator) and second neighbours (numerator):

\bea
\tan( \vartheta/2 )= \frac{ |\int \<\v r|\v A' \> \left[ \frac{p^2}{2m} + V_{\scriptsize \mbox{crystal}} \right] \< \v A| \v r \>  dV  | }{|\int \<\v r|\v B \> \left[ \frac{p^2}{2m} + V_{\scriptsize \mbox{crystal}} \right] \< \v A| \v r \>  dV  |}.
\label{3.3}
\eea
With these results, our wave functions produce an effective Dirac equation with mass describing the conduction band of monolayer MoS$_2$ in a very specific region of $k$ space, as reported in \cite{zahid}: between the points M and G there is the K point of maximal approach between the upper and lower bands, providing a reasonable region of $0.1$ for both adimensional wave numbers $k_x, k_y$. We may work with the experimental value $E_{\scriptsize \mbox{gap}}\approx 1.90$eV and the estimate $E_{\scriptsize \mbox{bandwidth}} \approx 3.25$eV in the region of interest. In fig. \ref{fig:VI}, we compare the numerically obtained dispersion relation using seven orbitals \cite{mak2010, kadantsev2012} with our initial relation (\ref{2.3.1}).   

\section{Conclusion \label{conclusion}}

We have shown that the stability group of the Dirac algebra defined on linear and hexagonal lattices can be applied to obtain new structures with the same propagation properties guaranteed by an invariant dispersion relation. Lorentz transformations have been included in a $\ppcal \tcal$-symmetric case that touches the field of artificial solids. Our efforts have also led to a simplified model of atomic orbitals in MoS$_2$, producing a deformed tight-binding model with second neighbours and matching the features of observed semiconduction bands. In connection with negative couplings and artificial realizations, e.g. microwaves in dielectric cylinders, it has been shown recently \cite{tony2016} that such opposite signs can be achieved by adding more structure to the arrays, giving rise to an additional Berry phase.

As an additional remark, we would like to point out that an infinite number of tight binding models with the same dispersion relation can be reached through the application of more general unitary transformations. For instance, in a two-dimensional lattice of $N^2$ sites, the elements of U$(N^2)$ gather all crystalline structures into the same equivalence class of dimensionality $N^4$. However, this suggests myriads of isospectral models which may not appear naturally in solids, nor in artificial constructions with resonators. Here, the use of quantum graphs seems more appropriate, but lies beyond the scope of the present paper.

Finally, it is worth noting that restricting ourselves to the stability group of Dirac matrices has led us to a simple description of gently deformed honeycomb crystals.

\ack

We are grateful to CONACyT for financial support under project CB2012-180585. YHE also wishes to thank CONACyT for {\it beca-cr\'edito\ }294863.

\section*{References}

\providecommand{\newblock}{}

\end{document}